# Observation of magnetoelectric effects in a $S=½$ frustrated spin chain magnet SrCuTe$_2$O$_6$

B. Koteswararao,[1,*] Kyongjun Yoo,[1,**] F. C. Chou,[2] and Kee Hoon Kim[1,a]

[1]*CeNSCMR, Department of Physics and Astronomy and Institute of Applied Physics, Seoul National University, Seoul 51-747, South Korea*

[2]*Center of Condensed Matter Sciences, National Taiwan University, Taipei 10617, Taiwan*

(Email: khkim@phya.snu.ac.kr)

The magnetoelectric (ME) effects are investigated in a cubic compound SrCuTe$_2$O$_6$, in which uniform Cu$^{2+}$ ($S=1/2$) spin chains with considerable spin frustration exhibit a concomitant antiferromagnetic transition and dielectric constant peak at $T_N \approx 5.5$ K. Pyroelectric $J_p(T)$ and magnetoelectric current $J_{ME}(H)$ measurements in the presence of a bias electric field are used to reveal that SrCuTe$_2$O$_6$ shows clear variations of $J_p(T)$ across $T_N$ at constant magnetic fields. Furthermore, isothermal measurements of $J_{ME}(H)$ also develop clear peaks at finite magnetic fields, of which traces are consistent with the spin-flop transitions observed in the magnetization studies. As a result, the anomalies observed in $J_p(T)$ and $J_{ME}(H)$ curves well match with the field-temperature phase diagram constructed from magnetization and dielectric constant measurements, demonstrating that SrCuTe$_2$O$_6$ is a new magnetoelectric compound with $S=1/2$ spin chains.

For more than a decade, the study of emergent multiferroic and magnetoelectric materials has been an active area of research due to fundamental interests on the spin-lattice coupling in frustrated magnets as well as application potential in low-power electronic devices [1-4]. Numerous multiferroic and magnetoelectric compounds have been found so far and studied extensively to understand the nature of spin ground states [1-4]. Among them, magnetoelectric/multiferroic materials with $S=1/2$ spins (mostly Cu$^{2+}$ ions) are of particular interest as they offer an opportunity of studying possible quantum effects in the vicinity of a new spin ground state through the dielectric measurements [5].

A few compounds with quasi-one-dimensional Cu$^{2+}$ spin ($S=1/2$) chains have been identified to become multiferroic; LiCuVO$_4$ [6, 7], LiCu$_2$O$_2$ [8-10], CuCl$_2$ [11], CuBr$_2$ [12], PbCuSO$_4$(OH)$_2$ [13, 14], and CuCrO$_4$ [15]. These magnets have a common structure of edge-sharing CuO$_4$ square planes, called CuO$_2$ ribbon chains. In such systems, the nearest-neighbor (*nn*) exchange interaction $J_1$ through the Cu-O-Cu paths is ferromagnetic ($J_1 < 0$), because the Cu-O-Cu angle is close to 90°. On the other hand, the next-nearest-neighbor (*nnn*) exchange interaction $J_2$ through the Cu-O-O-Cu exchange paths is antiferromagnetic ($J_2 > 0$), thus resulting in a non-trivial spin ground state due to the intra-chain competition between $J_1$ and $J_2$ interactions. It has been theoretically predicted that the spiral (or helical) spin structure, which is stabilized for $|J_2/J_1| > 1/4$ in the Heisenberg spin chains, can induce the ferroelectricity [16-20]. The spontaneous electric polarization (*P*) then satisfies the relation $P \propto Q \times e_3$ [16-18], where $Q$ and $e_3$ are the modulation vector and spin rotation axis of the spiral spin order, respectively. As a result of competition (spin frustration), a nontrivial spin ordering from a paramagnetic (paraelectric) to antiferromagnetic (ferroelectric) phase often occurs at the Neel temperature ($T_N$) of those materials; $LiCuVO_4$, $LiCu_2O_2$, and $PbCuSO_4(OH)_2$ indeed showed the spontaneous helical magnetic order and *P* to become multiferroic materials. Moreover, the exotic quantum phases such as spin-nematic, multi-polar order, and chiral spin order have been theoretically predicted [21–23]. For example, recent nuclear magnetic resonance (NMR) measurements on $LiVCuO_4$ concludes that the spin-nematic phase, if it does exist, can be found in a narrow field region at its saturated magnetic moments [24]. It was also found recently that dielectric response induced by short-range helimagnetic order exists in another spin chain system $Rb_2Cu_2Mo_3O_{12}$ [25-27], in which strong quantum spin fluctuation and low dimensionality is responsible for the short-range order [26]. However, it did not exhibit the electric polarization at zero field, but only appeared under applied magnetic fields [27].

On the other hand, identification of new multiferroic/magnetoelectric compounds with $S=1/2$ spins has become experimentally challenging. It is partly because their magnetoelectric effects are often too weak to detect electrically due to isotropic nature of $S=1/2$ spins and absence of strong spin-orbit coupling. Hence, polycrystalline samples often produced quite weak pyroelectric currents so that the intrinsic electrical signals were screened by leakage currents and free charges trapped in the grain boundaries after an electric poling [28]. Moreover, the scarcity of single crystals in the early stage of research made it difficult to investigate intrinsic ferroelectric polarization. On the other hand, it has been recently found in the study of a polycrystalline $TbMnO_3$ sample [29] that current measurements under a DC electric field bias is useful to identify intrinsic pyro- or magneto-current signals originating from the depolarization of ferroelectric polarization by mitigating the effect of the internal electric fields created by trapped charges in the grain boundaries.

$SrCuTe_2O_6$ is a new $Cu^{2+}$ based magnet, which has antiferromagnetic $S=1/2$ spin chains with frustrated antiferromagnetic inter-chain couplings [30, 31]. The presence of competing inter-chain coupling ($J_2$) is distinguished from the above mentioned ribbon chain materials. Here, we report the observation of intrinsic ferroelectric polarization by measurements of pyroelectric and magnetoelectric currents in the presence of a DC electric field bias. It is found that the magnetoelectric coupling becomes maximum at 4 K and 6 tesla. Moreover, the observed peak positions in the pyroelectric and magnetoelectric currents follow the previous magnetic phase diagram constructed from magnetization, heat capacity and dielectric constant measurements [31]. Our results point out that $SrCuTe_2O_6$ is a new magnetoelectric compound with a main antiferromagnetic interaction in the $S=1/2$ spin chain and the frustrated inter-chain interaction.

We first prepared the precursor $SrTe_2O_5$ by firing the pellets made of the stoichiometric amounts of $SrCO_3$ (99.99%) and $TeO_2$ (99.9%) at 450º C for 24 hours in the air, followed by one intermediate grinding. In the second step, the obtained $SrTe_2O_5$ sample was mixed with CuO in a molar ratio of 1:1. The pelletized mixture was sealed in an evacuated quartz tube and fired at 650ºC for 24 hours with two intermediate grindings. In order to avoid the $TeO_2$ evaporation, we have heated the pellets with a slow rate of 1°C per min. The powder x-ray diffraction measurements were performed at room temperature to confirm the absence of impurity phases. For measurements of pyroelectric and magnetoelectric currents, thin pellet forms with typical thickness ~ 1-2 mm were made during the synthesis and the pellet was further polished to make a thin plate with thickness ~0.4 mm. The polished pellet area is about 50.28 mm$^2$. Electrical contacts were made on both sides of the pellet by attaching thin copper wires (~85 μm diameter) with silver epoxy. Pyroelectric current measurements were performed using KE617 electrometer and inside PPMS$^{TM}$ (Quantum Design), which allows the measurements as a function of magnetic field (0 tesla - 9 tesla) and temperature (2- 300 K). For poling the sample, we have applied 2.5 kV/cm of electric field and 5.5 tesla of magnetic field at 10 K. Then, we have cooled down the specimen to lowers temperatures before starting electrical measurements upon warming at constant magnetic fields (pyroelectric currents) or upon ramping the magnetic field in the isothermal condition (magnetoelectric currents). The magnetization measurements were performed in the temperature-range from 2 to 5 K and in the magnetic field ($H$) range 0 - 7 tesla with a SQUID$^{TM}$ (Quantum design).

Figure 1 (a) shows the crystal structure of SrCuTe$_2$O$_6$, which comprises three one-dimensional (1D) $S$=1/2 (Cu$^{2+}$) uniform chains passing along *a, b,* and *c*-axes, respectively. The spin chains are formed by twisted plates of CuO$_4$ coupled via the O-O path with distance of 2.79 Å. The bond-distance of intra-chain Cu$^{2+}$ ions is 6.294 Å. The space group of this cubic system is $P4_132$ with a lattice constant $a$ =12.463 Å and $Z$=8 [32]. The main exchange interaction between Cu$^{2+}$ ($S$=1/2) ions in this spin chain was reported to be antiferromagnetic (AFM) with a magnitude of $J_1$/k$_B$≈−45 K and with considerable frustrated AFM inter-chain coupling ($J_2$/k$_B$) [28, 29]. As displayed in Fig. 1(b), the inter-chain couplings are in the form of hyper-Kagome network as in Na$_4$Ir$_3$O$_8$ [33] and PbCuTe$_2$O$_6$ [34], which is essential to provide the magnetic frustration between the Cu$^{2+}$ spins.

We first carried out the pyroelectric current $J_p$ measurements in a conventional way by poling the sample under electric field of $E_{pol}$ = +2.5 kV/cm and reached to a low temperature 2 K. When we removed $E_{pol}$, shorted out two electrodes, and measured $J_p(T)$ upon warming, we observed very weak pyroelectric current $J_p(T)$ signals. The measured $J_p(T)$ at $H$=5.5 tesla is shown in the inset (i) of Figure 2(a). There is a clear sign change in the $J_p(T)$ data after reversing the direction of poling electric field from $E_{pol}$= +2.5 kV/cm to -2.5 kV/cm. It is noted that a maximum value of $J_p(T)$ was obtained at $H$= 5.5 tesla. For other magnetic fields, the values of $J_p(T)$ were much smaller so that it was difficult to identify intrinsic signals. This observation still suggests that a weak polarization might exist in the sample under finite magnetic fields.

In order to find a more clear evidence of observing intrinsic $J_p(T)$ due to ferroelectric polarization, we have applied bias electric fields $E_{meas}$ = +2.5 kV/cm and -2.5 kV/cm during the measurements at $H$ = 6 tesla, as shown in inset (ii) of Figure 2(b). We first observe a similar sign change in the $J_p(T)$ data for reversing the direction of $E_{meas}$. Although the sign reversal effect is similar, the values of $J_p$ are nearly 8 times bigger than the $J_p$ without $E_{meas}$. Furthermore, the $J_p$ in the presence of $E_{meas}$ clearly exhibits a peak followed by dip structure, indicating the crossing of two temperature-induced phase transitions at $H$ = 6 tesla (see, Fig. 6 below). In recent pyroelectric measurements in a polycrystalline TbMnO$_3$, the spurious $J_p$ due to internal electric field coming from the high temperature poling procedure can mimic the intrinsic $J_p$ coming from the intrinsic depolarization current but with a different sign. As a result, the spurious $J_p$ could indeed reduce or screen the intrinsic $J_p$ data. Furthermore, the study indicated that the application of small electric field bias during $J_p$ can result in the intrinsic signals. From the enhancement of $J_p$ in the presence of $E_{meas}$, we presume that the similar effect could also exist in our polycrystalline specimen. Therefore, we have applied $E_{meas}$ always in the rest of the measurements.

The measured data of $J_p(T)$ for various magnetic fields are shown in the inset (iii) of Figure 2(b), which might contain the contributions of both intrinsic depolarization/polarization currents and leakage currents. In order to extract the intrinsic $J_p$ generated under magnetic fields without being subject to leakage currents of the specimen, we have subtracted the data of $J_p(T)$ measured at $H=0$ tesla from that of $J_p(T)$ measured at various $H$ from 1 to 7 tesla (see Figure 2(a) and 2(b)). At $H=0$ tesla, we found that $J_p(T)$ becomes almost zero across $T_N$ even under the bias electric field. The enhancement of $J_p(T)$ at low temperatures is certainly due to the application of magnetic field. This observation suggests that $SrCuTe_2O_6$ is not a multiferroic but a magnetoelectric compound.

Figure 3 summarizes the change in the electric polarization ($\Delta P$) as obtained from integration of the pyroelectric currents with respect to time. The calculated $\Delta P$ values were found to be quite small as compared to the Cu-based multiferroics as well as other multiferroics with ferroelectricity driven by magnetic order [6-15]. Interestingly, the ferroelectric transition temperatures, at which $\Delta P$ appears (indicated by dashed arrows in Figure 3), are in good agreement with the corresponding data obtained from the magnetization and dielectric constant data [30, 31]. The magnitude of $\Delta P$ is also varied with the magnetic fields. For instance, the large values of $\Delta P$ are observed for fields of $5.5 \leq H \leq 6$ tesla. The $H$-dependent behavior of $\Delta P$ values and $H$-dependent ferroelectric transitions establish that $SrCuTe_2O_6$ is a magnetoelectric (ME) compound because $\Delta P$ develops across $T_N$ under a finite magnetic field.

In order to confirm further the ME behavior, we measured the isothermal ME current $J_{ME}(H)$ (see Figure 4). At 4 K, $J_{ME}(H)$ data clearly exhibit a peak at about 6 tesla. Moreover, upon changing the polarity of $E_{meas}$, $J_{ME}(H)$ shows the sign reversal, revealing the ferroelectric domain change as an origin of the $J_{ME}(H)$ peaks. The sign reversal can be also observed in the $H$-increasing and $H$-decreasing runs (the inset of Figure 4), constituting another evidence that $J_{ME}(H)$ comes from the $H$-induced variation of ferroelectric polarization. A similar sign reversal was previously observed in multiferroic $NaFeGe_2O_6$ polycrystalline samples [35], in which the multiferroicity and the ME effects were later confirmed in the single crystals as well [36].

To get insights on the origin of the peaks in the $J_{ME}$ data, we measured $J_{ME}(H)$ at various temperatures from 2 to 5 K. We then calculated $\Delta P(H)$ and its derivative ($dP/dH$) (see Figure 5(a) and (b)). Here, we note that the $dP/dH$ plot can show the ME coupling in the units of ps/m while $dP/dH$ is essentially same as $J_{ME}(H)$. As seen in Figure 5(a), $\Delta P(H)$ curves exhibit strong field and temperature dependence. At 2 K, the $\Delta P(H)$ data develop peaks at $H=\sim 4$ and $\sim 6$ tesla, and those fields showing the peaks in the $\Delta P(H)$ curves change with temperature. The systematic shifts of the $\Delta P(H)$ peaks at different temperatures can also be identified as sharp dips in the $dP/dH$ curves in Figure 5(b). It is interesting to find that the systematic shifts of the dips in the $dP/dH$ curves are quite similar to the peak shifts in the magnetic data, *i.e.*, $dM/dH$ versus $H$ curves in Figure 5(c). Those peaks in the $dM/dH$ curves stem from the spin-flop transitions. Therefore, the changes of electric polarization are directly coupled to the spin-flop transitions as expected in the magnetoelectric materials. Similar peaks in the $dP/dH$ curves associated with the spin-flop transitions have been also observed in several magnetoelectric polar magnets such as $Ni_3TeO_6$ [37] and $Fe_2Mo_3O_{12}$ [38]. It is also noteworthy that the maximum of the absolute ME coefficient $|dP/dH|^{max}$ increases almost linearly with temperature and develops a maximum of 7 ps/m at 4 K and 6 tesla (see, inset of Figure 5(b)). The $|dP/dH|^{max}$ subsequently decreases steeply at higher temperatures. The physical origin of the increase in the $|dP/dH|^{max}$ in the vicinity of a bicritical point involving the spin-flop transition might be worthwhile to investigate in future studies.

All of the results in Figure 5 unambiguously support the existence of ME coupling in this frustrated spin chain material. The transition temperatures observed in the $\Delta P(T)$ curves at various $H$'s (as peaks in $J_p(T)$ curves) and the peak positions in the $\Delta P(H)$ curves at various $T$'s are plotted in Figure 6. Interestingly, the phase boundaries extracted from those electrical measurements are in good agreement with the reported phase-diagram constructed by magnetic, heat capacity, and dielectric constant results [31].

The ferroelectric transition is often coined to the magnetic transition in many spin chain materials, which is intimately linked to the presence of spin frustration and spin-lattice coupling. There are many examples of ribbon chain materials as summarized in table 1, most of which have the intra-chain competition between FM $nn$ and AFM $nnn$ couplings. The present spin chain material $SrCuTe_2O_6$ is rather different from those of existing ribbon chain multiferroics because the frustration mostly comes from the AFM inter-chain coupling ($J_2$) with the strength $zJ_2/J_1 \approx 0.1$, where z is the coordination number, denoting the number of inter-chain couplings [31]. On the other hand, the space group $P4_132$ of this material is polar so that without spiral or helical spin order, it might allow the electric polarization through the exchange-striction mechanism under finite magnetic fields. While it can in principle become ferroelectric even at zero magnetic field through the exchange striction mechanism, our data do not support the scenario within the resolution of pyroelectric current measurements. Yet, the origin of ME coupling could be better understood when the spin ordering pattern is revealed by e.g. neutron diffraction studies.

Regardless of the origin of its ME coupling, the value of $\Delta P$ induced under magnetic fields is quite small. Several reasons may explain this. First, the main exchange path involves Cu-O-O-Cu with rather a long Cu-Cu bond distance (6.294 Å) as compared with the Cu-Cu bond distances (~2.9 Å) of ferromagnetic ribbon chain compounds, which will be certainly detrimental to induce large electric polarization. Second, the chains are passing all three directions, rather than along a particular direction as in other spin chain systems. The resultant polarization might be then minimized due to the average effects along the three chain contributions. Moreover, the present system has the cubic structure, which usually does not favor the large electric polarization and ME coupling, unlike the low-symmetry structures. Third, as explained in Table 1, the frustration comes from the inter-chain couplings of an AFM-type, which usually have more quantum fluctuations than that of ferromagnetic ribbon chains. For example, when the frustrated intra-chain coupling (nnn interaction) was FM , the corresponding material induced a spiral/helical LRO ground state, while the frustrated AFM intra-chain coupling opened a robust, disordered spin gap due to the enhanced fluctuations [44]. Meanwhile, there is another kind of system $Rb_2Cu_2Mo_3O_{12}$ , which has significant frustration and large quantum fluctuations to induce short-range order albeit it has the frustrated AFM chain coupling (see Table 1). This system does not develop any electric polarization at $H=0$ tesla, despite having a peak in the dielectric constant. However, it exhibits electric polarization at finite magnetic fields [27]. The small ΔP induced by magnetic field in SrCuTe$_2$O$_6$ is quite similar to the phenomena observed in Rb$_2$Cu$_2$Mo$_3$O$_{12}$. Thus, SrCuTe$_2$O$_6$ might also have strong fluctuation coming from low-dimensionality of the AFM intra-chain and the frustrated inter-chain AFM couplings. Therefore, the smallness of ΔP and the weak ME coupling could represent the intrinsic feature of enhanced spin fluctuation in this system.

In conclusion, we have found magnetoelectric behavior in a frustrated, antiferromagnetic spin chain compound SrCuTe$_2$O$_6$ below its magnetic transition by use of the pyroelectric $J_p(T)$ and magnetoelectric $J_{ME}(H)$ current measurements under a constant electric field bias. The observed magnetoelectric coupling coefficients (d$P$/d$H$) versus $H$ data follow exactly similar behavior observed in the data of d$M$/d$H$ vs $H$ curves, constituting a clear evidence for the magnetoelectric coupling. The transition temperatures or magnetic fields found in $J_p(T)$ and $J_{ME}(H)$ data follow the previous phase-diagram based on magnetization, heat capacity and dielectric constant measurements [31]. The presence of strong quantum fluctuations in this quasi-one-dimensional spin chain material could be a reason for having small electric polarization. Detailed neutron diffraction, NMR, µSR measurements would be useful for understanding further the origin of magnetoelectric coupling and fluctuating spin ground states.

**Acknowledgements**

This work at SNU has been supported by the National Creative Research Initiative (2010-0018300). BK thanks the partial support of DST INSPIRE faculty award-2014 scheme. FCC acknowledges the support from the Ministry of Science and Technology in Taiwan under project number MOST-102-2119-M-002-004.

(* represents the present afflation as *School of Physics, University of Hyderabad, Central University PO, Hyderabad 500046, India*,** Equal contributions as the first authors, + a corresponding author)

**Figures and captions**

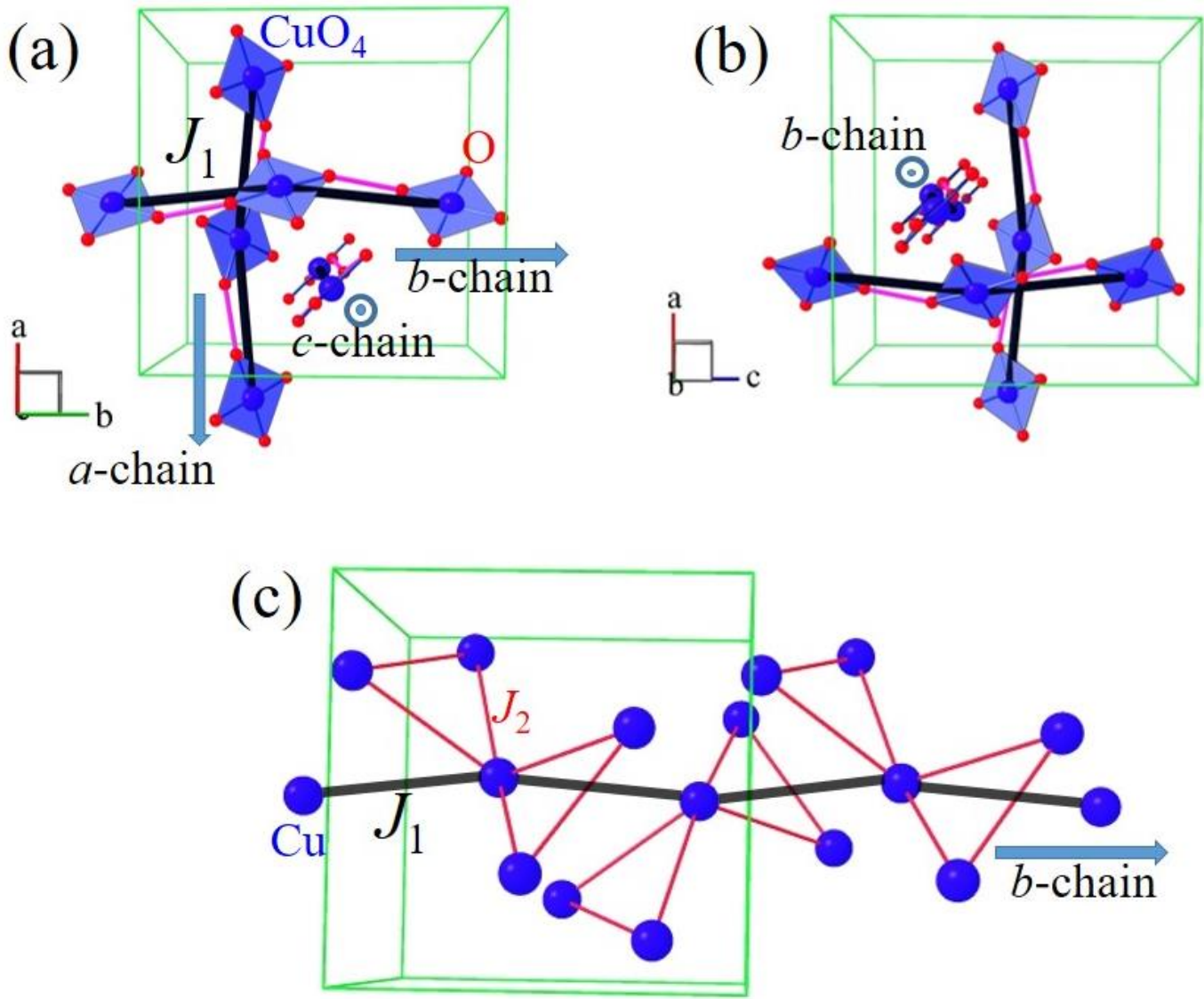

Figure 1(a) & (b). The crystal structure of SrCuTe$_2$O$_6$ showing uniform one-dimensional Cu$^{2+}$ ($S$=1/2) spin chains along all the crystallographic $a$-, $b$-, and $c$-directions. (c) Spin chain with intra-chain ($J_1$) and frustrated inter-chain ($J_2$) interactions. For simplicity, we have shown only the chain along $b$-direction.

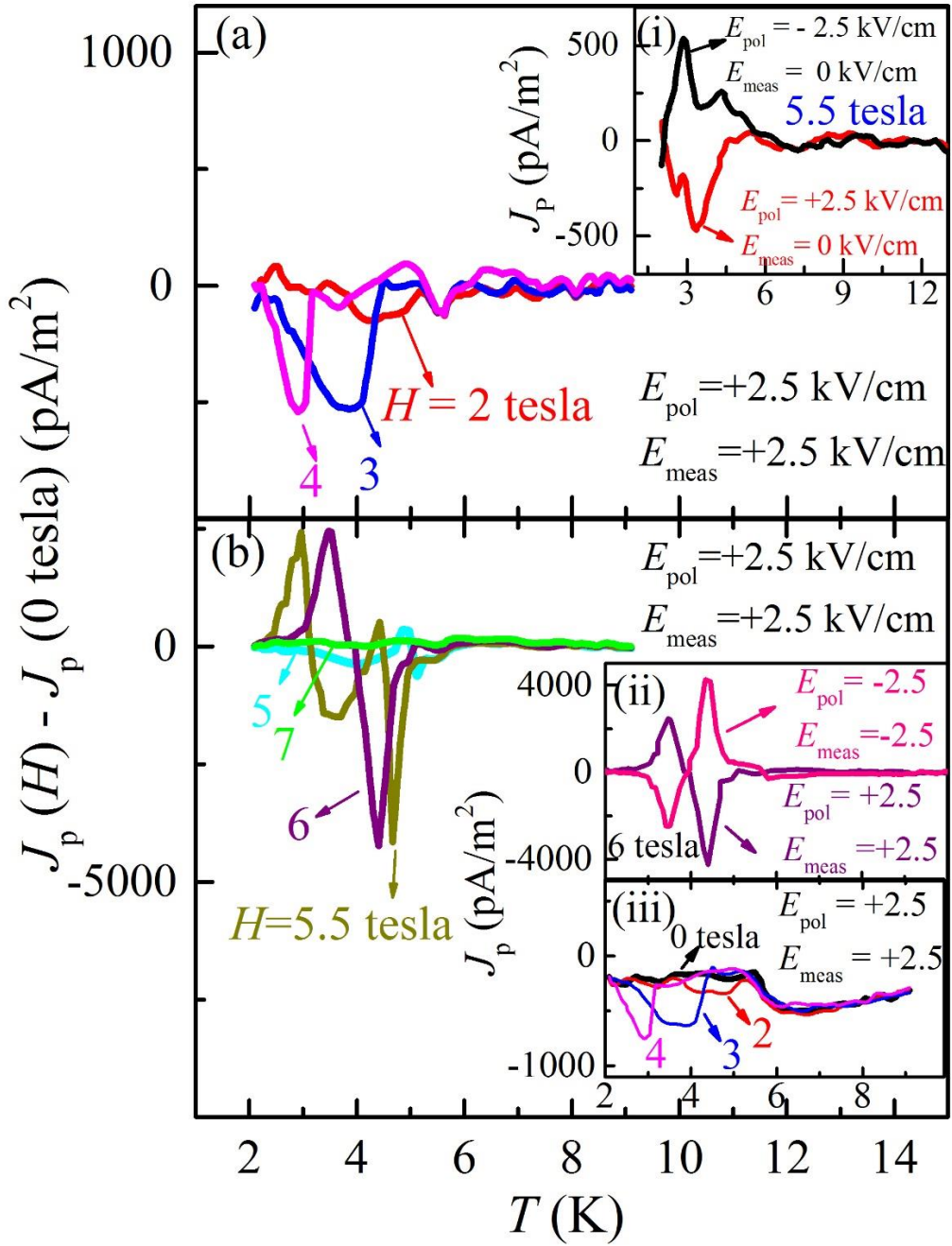

Figure 2 (a) & (b) Pyroelectric currents at various constant magnetic fields ($\mu_0 H$ = 1 to 7 tesla), which are obtained after the subtraction of the same data at $H=0$ tesla ($J_p(H)-J(0\ \text{tesla})$). The inset (i) shows $J_p(T)$ data at $H$ =5.5 tesla measured upon warming without electric field bias $E_{meas}$ after cooling down the sample under electric field poling ($E_{pol}$ = +2.5 kV/cm and -2.5 kV/cm). The inset (ii) shows the same $J_p(T)$ data at $H$ = 6 tesla measured at $E_{meas}$ = 2.5 kV/cm after cooling down the sample under $E_{pol}$. The inset (iii) displays the measured $J_p$ data under $E_{meas}$ at various magnetic fields.

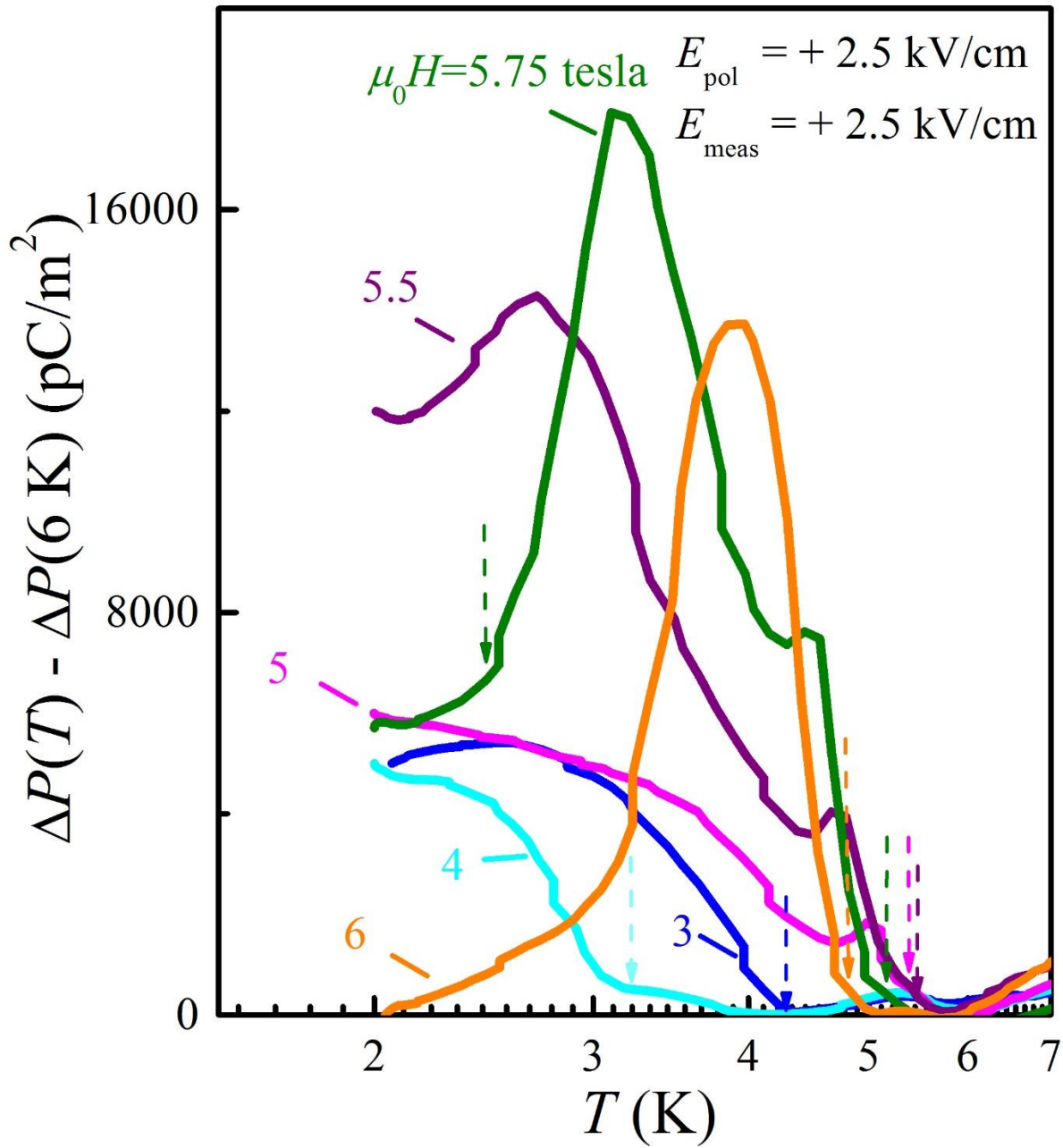

Figure 3 Change in the electric polarization ΔP, after subtracting the value of ΔP at 6 K, as a function of T at various magnetic fields ranging from 3 to 6 tesla. The dashed arrow marks indicate the ferroelectric transitions where the ΔP has a sudden change, which are also summarized in Fig. 6.

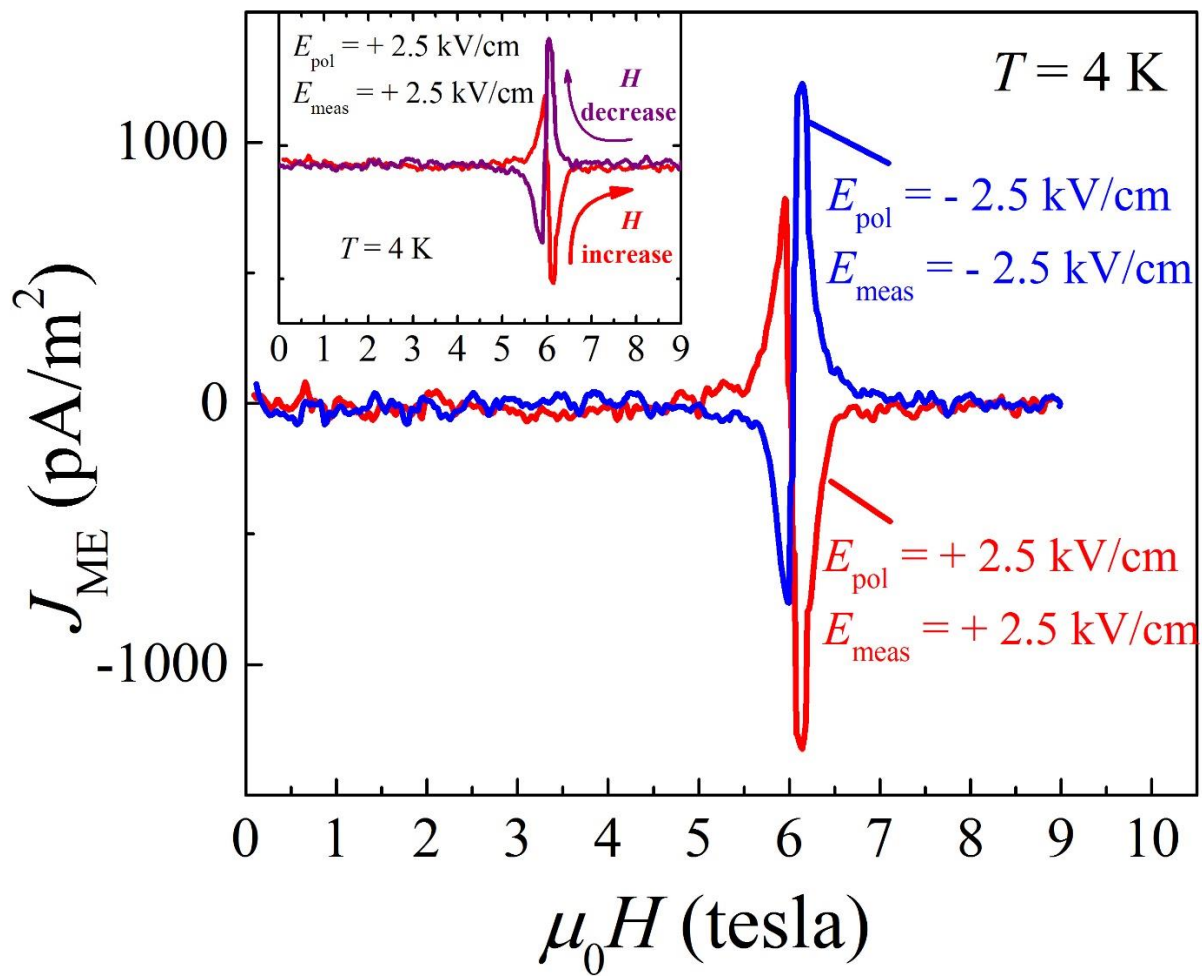

Figure 4 Magnetoelectric current $J_{ME}$ at 4 K measured in the presence of a constant bias electric field ($E_{meas}$). Inset show the plot of $J_{ME}$ measured in the $H$-sweeping up and $H$-sweeping down runs.

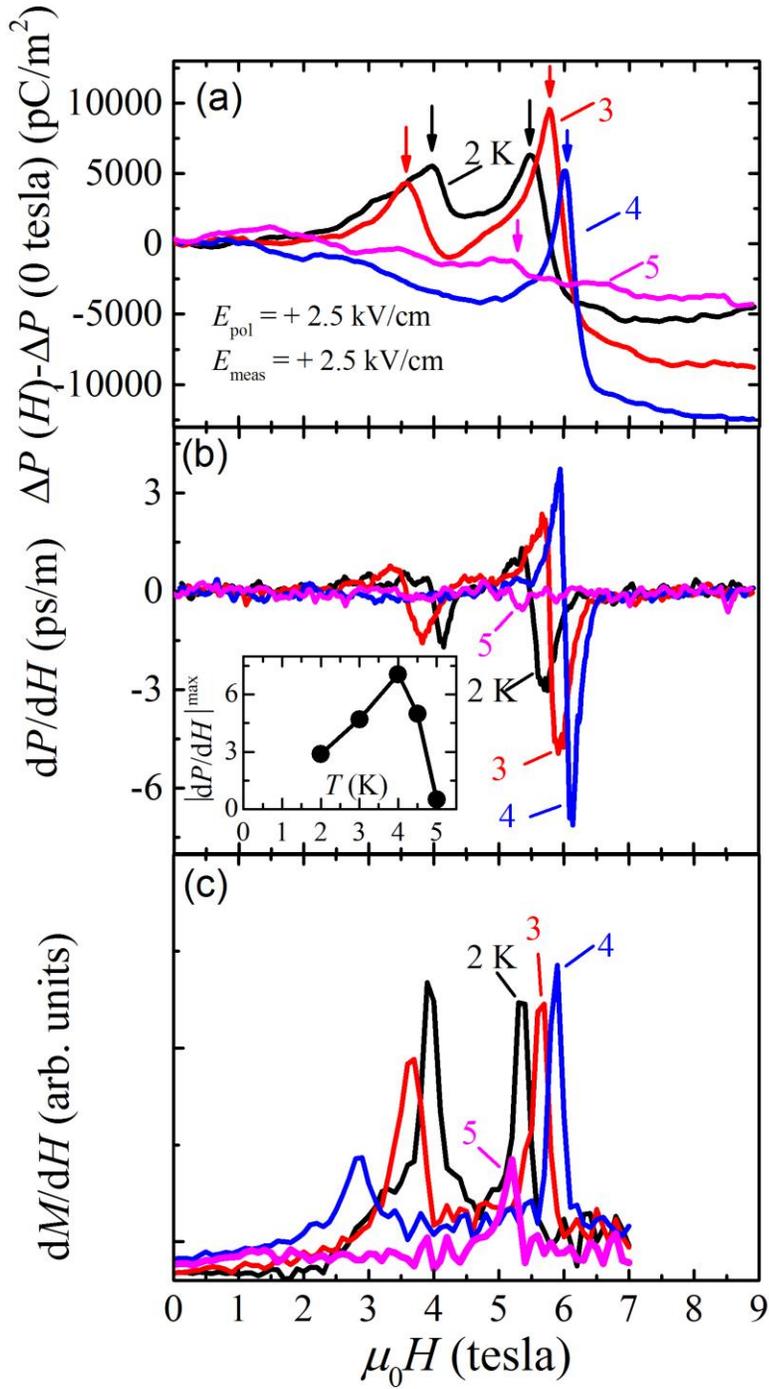

Figure 5 (a) Change in the electric polarization ΔP as a function of magnetic field ($\mu_0H$) for various temperatures from 2 to 5 K. The arrow marks indicate the changes in the polarization. (b) $H$-derivative of electric polarization (d$P$/d$H$) in the presence of a constant electric field $E_{meas}$ = +2.5 kV/cm at various temperatures 2, 3, 4, and 5 K. Inset shows the plot of the maximum values of d$P$/d$H$ versus $T$. (c) $H$-derivative of magnetization (d$M$/d$H$) versus $\mu_0H$.

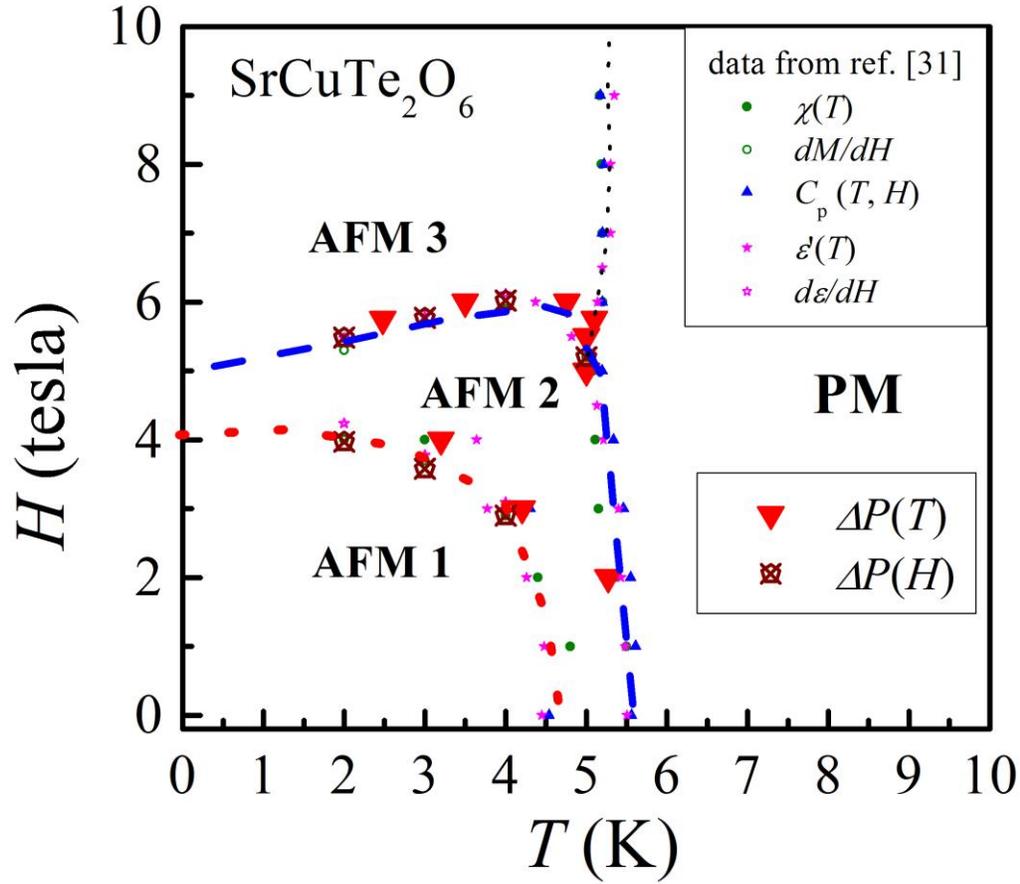

Figure 6 The phase diagram was constructed from the observed anomalies in the data of $\Delta P(T)$ at various fields and the isothermal $\Delta P(H)$ data at various temperatures. Previous phase boundaries of different antiferromagnetic (AFM) regions from the Ref. [31], which were determined from magnetic susceptibility $\chi(T)$, dielectric constant ($\varepsilon'$), and heat capacity $C_p$ results, are also overlapped. AFM and PM refer to 'antiferromagnetic' and 'paramagnetic' phases.

Table 1. Details of the spin ground states in several $S=1/2$ frustrated spin chain materials. LRO and SRO refer 'long range order' and 'short range order', while FM and AFM represent 'ferromagnetic' and 'antiferromagnetic'.

| Compound | $J_1$ (K) | $J_2$ (K) | Type of frustration | $T_N$ (K) | Magnetic order type | reference |
|---|---|---|---|---|---|---|
| $LiCu_2O_2$ | -69 (FM) | 73 (AFM) | Intra-chain | 24.6, 23.2 | Helical LRO | [8, 9, 39, 40] |
| $LiCuVO_4$ | -19 (FM) | 44 (AFM) | Intra-chain | 2.3 | Helical LRO | [6, 7, 41, 42] |
| $PbCu(SO_4)(OH)_2$ | -100 (FM) | 36 (AFM) | Intra-chain | 2.8 | Helical LRO | [13, 14, 43] |
| $Rb_2Cu_2(MoO_4)_3$ | -138 (FM) | 51 (AFM) | Intra-chain | 8 K (under $H$ up to 9 tesla) | Helical SRO | [25-27] |
| $SrCuTe_2O_6$ | 50 (AFM) | 10 (AFM) | Inter-chain | 5.5, 4.5 | | [30, 31] |